\documentclass[%
 reprint,
 amsmath,amssymb,
 aps,
]{revtex4-2}

\usepackage{graphicx}
\usepackage{dcolumn}
\usepackage{bm}
\usepackage[colorlinks=true]{hyperref}

\begin{document}

\preprint{APS/123-QED}

\title{Cooperative Suppression Strategy for Dual Thermal Transport Channels in Crystalline Materials}

\author{Yu Wu}
\email{wuyu@njnu.edu.cn}
\affiliation{Advanced Thermal Management Technology and Functional Materials Laboratory, Ministry of Education Key Laboratory of NSLSCS, School of Energy and Mechanical Engineering, Nanjing Normal University, Nanjing 210023, P. R. China}

\author{Shuming Zeng}
\affiliation{College of Physics Science and Technology, Yangzhou University, Jiangsu, Yangzhou 225009, China}

\author{Chenhan Liu}
\email{chenhanliu@njnu.edu.cn}
\affiliation{Advanced Thermal Management Technology and Functional Materials Laboratory, Ministry of Education Key Laboratory of NSLSCS, School of Energy and Mechanical Engineering, Nanjing Normal University, Nanjing 210023, P. R. China}

\author{Hao Zhang}
\affiliation{College of Future Information Technology and Department of Optical Science and Engineering and State Key Laboratory of Photovoltaic Science and Technology, Fudan University, Shanghai 200433, China}

\author{Liujiang Zhou}
\affiliation{School of Physics, State Key Laboratory of Electronic Thin Films and Integrated Devices, University of Electronic Science and Technology, Sichuan, Chengdu 610054, China}

\author{Ying Chen}
\affiliation{Eastern Institute of Technology, Zhejiang, Ningbo 315200, China}

\author{Su-Huai Wei}
\email{suhuaiwei@eitech.edu.cn}
\affiliation{Eastern Institute of Technology, Zhejiang, Ningbo 315200, China}

\begin{abstract}

We propose a novel design principle for achieving ultralow thermal conductivity in crystalline materials 
via a “heavy-light and soft-stiff” structural motif. 
By combining heavy and light atomic species with soft and stiff bonding networks, 
both particle-like ($\kappa_p$) and wave-like ($\kappa_c$) phonon transport channels are concurrently suppressed. 
First-principles calculations show that this architecture induces a hierarchical phonon spectrum: 
soft-bonded heavy atoms generate dense low-frequency modes that enhance scattering and reduce $\kappa_p$, 
while stiff-bonded light atoms produce sparse high-frequency optical branches 
that disrupt coherence and lower $\kappa_c$. 
High-throughput screening identifies Tl$_4$SiS$_4$ ($\kappa_p$ = 0.10, $\kappa_c$ = 0.06 W/mK) 
and Tl$_4$GeS$_4$ ($\kappa_p$ = 0.09, $\kappa_c$ = 0.06 W/mK) 
as representative candidates with strongly suppressed transport in both channels. 
A minimal 1D triatomic chain model further demonstrates the generality of this mechanism, 
offering a new paradigm for phonon engineering beyond the conventional $\kappa_p$–$\kappa_c$ trade-off.

\end{abstract}

\maketitle


Because phonon shows both particle and wave like features, in pursuit of finding crystalline materials with ultralow lattice thermal conductivity ($\kappa_L$), it faces a fundamental challenge: the intrinsic competition between 
particle-like ($\kappa_p$) and coherent wave-like ($\kappa_c$) phonon transport channels\cite{Zeng2024,Xiong2025,Li2024d,Zheng2024,Wu2023b,Ji2024}. 
While complex crystal structures can suppress $\kappa_p$ via enhanced phonon–phonon scattering, 
they often simultaneously promote $\kappa_c$ by generating dense phonon spectra 
and numerous near-degenerate phonon pairs, 
which enhances vibrational mode coupling and coherent tunneling\cite{Wu2024a,DiLucente2023,Simoncelli2019,Zhang2025,Zeng2025}. 
This antagonistic relationship creates a thermodynamic constraint that prevents the lattice thermal conductivity $\kappa_L$ in bulk crystals from falling below a certain threshold at room temperature\cite{Zeng2024}.

The fundamental challenge in achieving ultralow thermal conductivity in crystalline materials stems from the competing nature of phonon transport channels, with microscopic origins rooted in the fundamental parameters governing phonon transport. Within the Debye framework \cite{Terry2004a}, lattice thermal conductivity ($\kappa_L$) depends critically on heat capacity ($C_V$), phonon group velocity ($v_g$), and relaxation time ($\tau$)---all strongly mediated by bonding characteristics and structural complexity \cite{Wu2024a}. While established strategies like complex crystal structures \cite{Jia-Jun2019,Liu2022}, soft bonds, or anharmonic features (including rattling atoms \cite{Christensen2008,Tadano2015,Jana2017,Wu2025}, lone-pair electrons \cite{Nielsen2013,Skoug2011}, and resonant bonding \cite{Lee2014,Ji2022}) can suppress particle-like transport ($\kappa_p$) through enhanced scattering and reduced $v_g$, they invariably increase phonon mode density and coupling---thereby amplifying wave-like coherence ($\kappa_c$). However, the discovery of exceptional materials like $X_6$Re$_6$S$_8$I$_8$ ($X = \text{Rb, Cs}$) that circumvent this trade-off \cite{Xiong2025} revealed new design possibilities, though the underlying principles remained obscured by the complex interplay of atomic-scale interactions and collective vibrational phenomena.

Our work establishes a universal ``heavy-light, soft-stiff" design paradigm that systematically decouples these transport channels by engineering hierarchical phonon spectra. This approach combines precise mass contrast and bonding hierarchy to create materials where dense low-frequency modes from heavy, soft-bonded units suppress $\kappa_p$ through enhanced scattering, while sparse high-frequency branches from light, stiff-bonded components minimize $\kappa_c$ via vibrational decoherence. Validated through first-principles calculations and materials screening, this paradigm identifies promising candidates like Tl$_4$SiS$_4$ and Tl$_4$GeS$_4$ that achieve simultaneous suppression of both transport channels ($\kappa_{\text{L}} = 0.15$~W/mK), establishing a new pathway for phonon engineering beyond conventional limitations.


The calculations are performed using the Vienna Ab Initio Simulation Package (VASP) based on density functional theory (DFT)\cite{Kresse1996}, employing the projector augmented wave (PAW) method with the PBEsol exchange--correlation functional\cite{Perdew2008}. A plane-wave cutoff energy of 500 eV is used. Atomic positions are optimized with an energy convergence criterion of $10^{-5}$ eV between consecutive steps and a maximum Hellmann-Feynman force tolerance of $10^{-3}$ eV/\r{A}. For the ab initio molecular dynamics (AIMD) simulations, the supercell was constructed shown in Table~S1 with $\Gamma$-centered 1$\times$1$\times$1 $\mathbf{k}$-mesh. The simulations run for 30 ps with a 1 fs timestep. Interatomic force constants (IFCs) are extracted from AIMD trajectories using the temperature-dependent effective potential (TDEP) method\cite{Hellman2013}. The lattice thermal conductivity ($\kappa_L$) and related parameters, including phonon relaxation times, are computed using the ShengBTE software\cite{Li2014,Han2022}, which implements an iterative solution scheme. The $\mathbf{q}$-mesh shown in Table~S1 is adopted in the first irreducible Brillouin zone, with a Gaussian smearing width of 0.1. 

\begin{figure*}[ht!]
\centering
\includegraphics[width=1\linewidth]{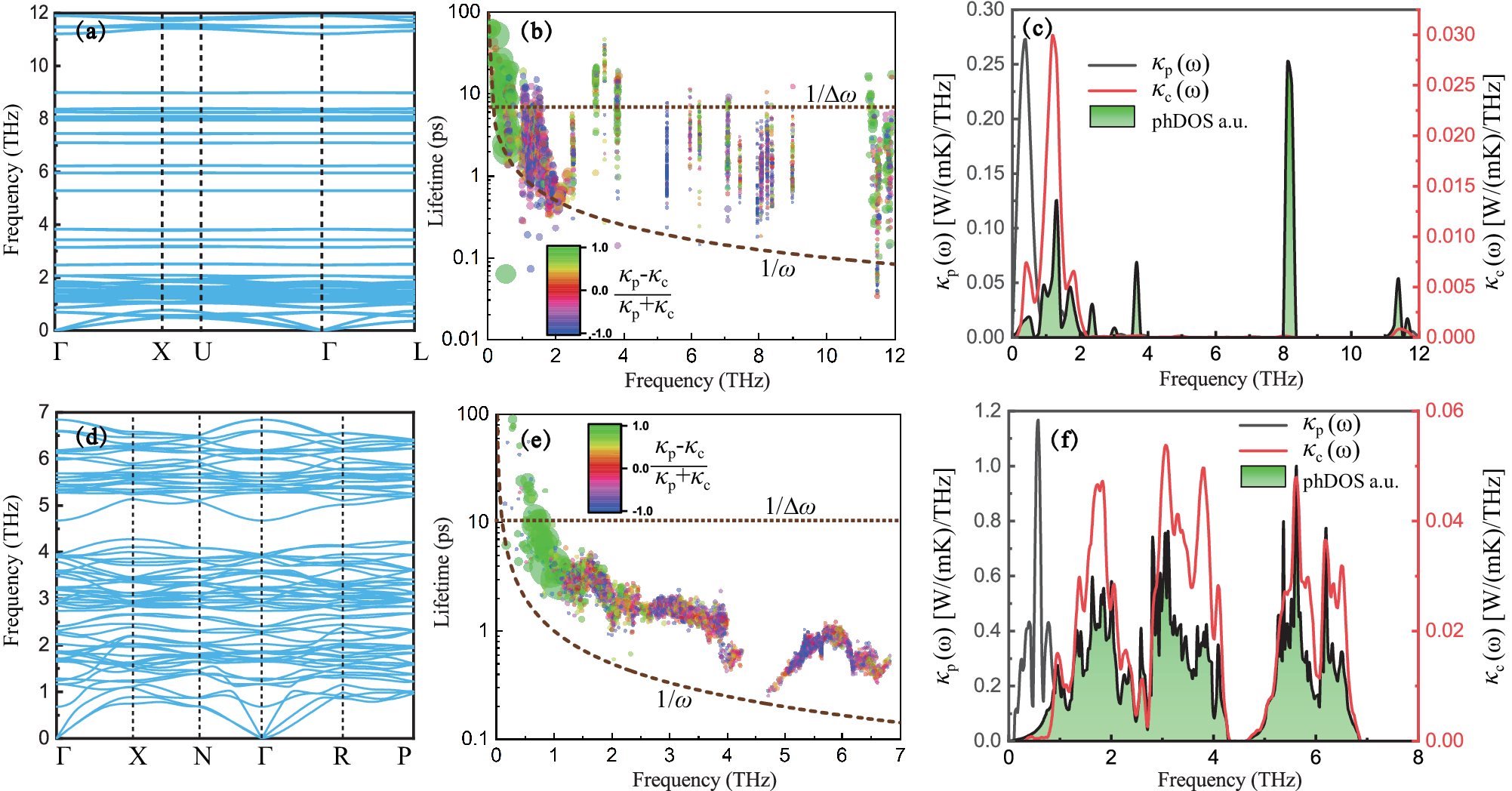}
\caption{(a, d) The phonon dispersion of Rb$_6$Re$_6$S$_8$I$_8$ and MgAgSb at 300 K. (b, e) Phonon lifetime of Rb$_6$Re$_6$S$_8$I$_8$ and MgAgSb versus frequency at 300 K, with scatter point areas proportional to their contribution to $\kappa_L$ (green: particle-like $\kappa_p$, blue: glass-like $\kappa_c$). The Wigner limit ($\tau_{\text{Wigner}} = 1/\Delta\omega$) and Ioffe-Regel limit ($\tau_{\text{Ioffe-Regel}} = 1/\omega$) are shown as dotted lines. (c, f) The differential of cumulative $\kappa_p$, $\kappa_c$ and dimensionless phDOS of Rb$_6$Re$_6$S$_8$I$_8$ and MgAgSb at 300 K.}
\label{Fig1}
\end{figure*}

The compound Rb$_6$Re$_6$S$_8$I$_8$ demonstrates an exceptional case of simultaneous suppression of both particle-like ($\kappa_p$=0.15 W/mK) and wave-like ($\kappa_c$=0.02 W/mK) thermal conductivity channels\cite{Xiong2025}. To understand the underlying mechanisms governing this unusual transport behavior, we analyze its phonon properties in detail. Figure~\ref{Fig1}(a) displays the phonon dispersion of Rb$_6$Re$_6$S$_8$I$_8$, showing distinct spectral characteristics: a dense concentration of phonon branches below 2 THz and sparse, exceptionally flat branches above 2 THz. The flat high-frequency branches create weak scattering channels because phonon interactions require strict energy and momentum conservation. While momentum conservation is easily satisfied in flat branches, energy conservation (e.g., for processes like $\omega + \omega_1 = \omega_2$) requires exact matching with existing phonon frequencies, making many scattering events forbidden. This leads to the long phonon lifetimes observed in Figure~\ref{Fig1}(b) for high-frequency modes. The sparse high-frequency spectrum also results in large frequency differences ($\omega-\omega'$) between phonon pairs, suppressing coherent transport contributions to $\kappa_c$.

Figure~\ref{Fig1}(c) reveals the fundamental competition between $\kappa_p$ and $\kappa_c$ suppression mechanisms in Rb$_6$Re$_6$S$_8$I$_8$. While dense low-frequency phonon branches (0-0.8 THz) with large group velocities provide sufficient scattering channels to reduce $\kappa_p$ through enhanced phonon-phonon interactions, the same phDOS would normally increase $\kappa_c$ by creating small frequency differences between phonon pairs. However, the material's unique spectral distribution, with sharply reduced phonon density above 2 THz, breaks this conventional competition. The differential $\kappa_c$ analysis shows that while $\kappa_c$ primarily originates from the 0.8-2 THz range where phonon branches are densely packed, high-frequency phonons ($>$2 THz) contribute minimally due to their sparse distribution, as confirmed by the correspondence with phonon density of states.

In contrast, MgAgSb ($\kappa_p$=0.47 W/mK, $\kappa_c$=0.13 W/mK) shows fundamentally different behavior in Figures \ref{Fig1}(d-f). Its uniformly dense and continuous phonon spectrum (Figure~\ref{Fig1}(d)) creates abundant scattering channels throughout the entire frequency range. While this does reduce $\kappa_p$ through increased scattering, it also leads to shorter phonon lifetimes (Figure~\ref{Fig1}(e)) and, more importantly, enables significant $\kappa_c$ contributions. Most phonons fall between the Wigner and Ioffe-Regel limits, and Figure~\ref{Fig1}(f) demonstrates that $\kappa_c$ receives substantial contributions across a broad frequency range (1-7 THz) due to the small frequency differences ($\omega-\omega'$) between phonon pairs enabled by the dense spectrum.

The key to Rb$_6$Re$_6$S$_8$I$_8$'s exceptional performance lies in its strategic spectral distribution. While conventional wisdom suggests that dense phonon spectra universally benefit thermal conductivity reduction, our analysis reveals that only low-frequency density (below 2 THz) is essential for $\kappa_p$ suppression through enhanced scattering. The additional high-frequency density in materials like MgAgSb provides diminishing returns for $\kappa_p$ reduction while substantially enhancing $\kappa_c$ through increased phonon coherence. Rb$_6$Re$_6$S$_8$I$_8$ achieves optimal decoupling by concentrating phonon density where it matters most (low frequencies) while maintaining spectral sparsity at higher frequencies where it would unnecessarily boost $\kappa_c$. This innovative approach to phonon band engineering successfully resolves the traditional competition between $\kappa_p$ and $\kappa_c$ suppression mechanisms, providing a new design principle for ultralow-$\kappa$ materials.

%
%

\begin{figure*}[ht!]
\centering
\includegraphics[width=1\linewidth]{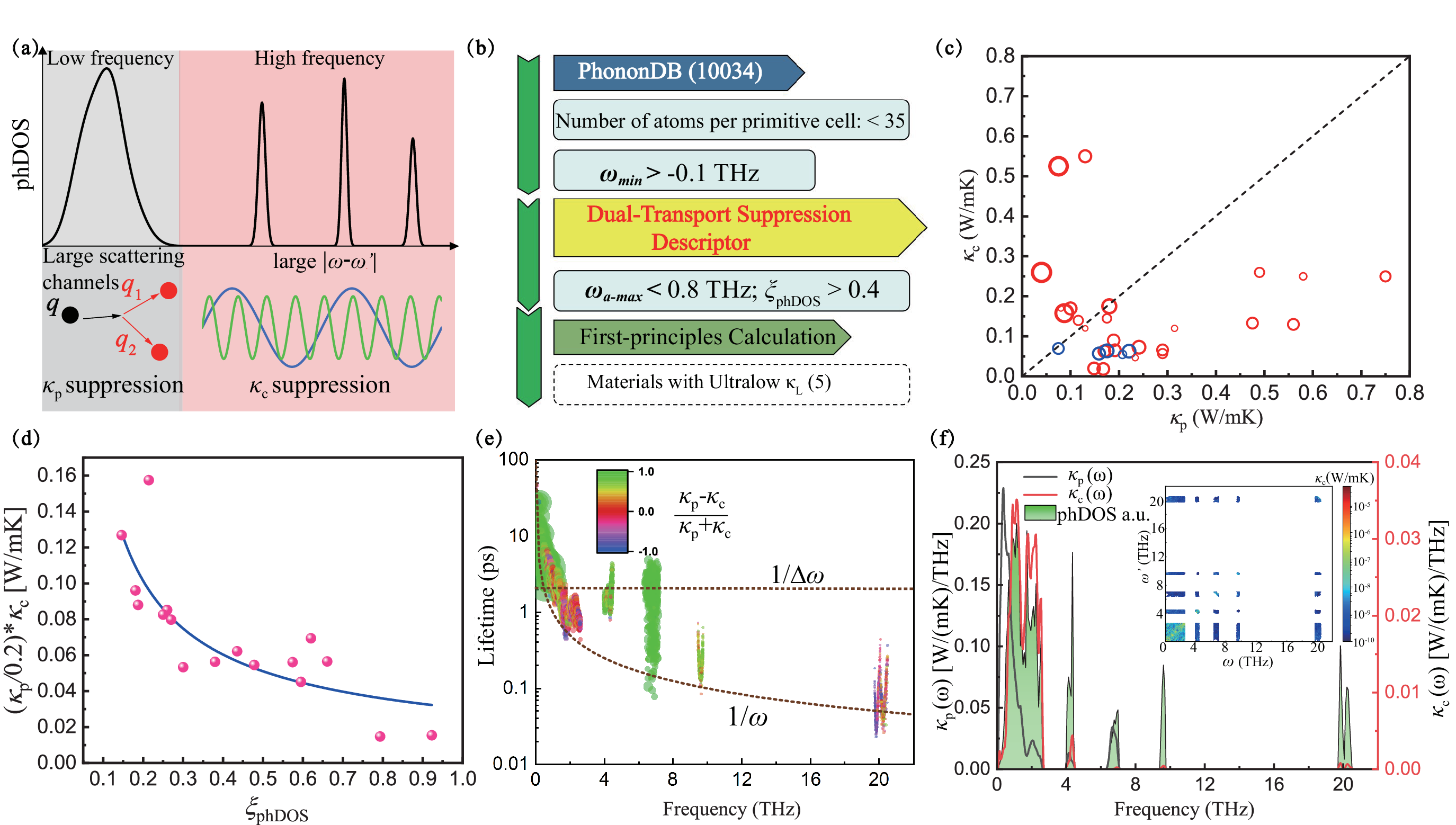}
\caption{(a) Characteristic phDOS profile for materials enabling simultaneous $\kappa_p$-$\kappa_c$ suppression, featuring a broad low-frequency peak and sparse high-frequency peaks. (b) High-throughput screening workflow applied to the PhonoDB database. (c)Room-temperature $\kappa_p$-$\kappa_c$ values with circle sizes representing primitive cell atom counts, showing screened candidates (blue) compared to reference materials (red)\cite{Xiong2025,Zhang2025,Zeng2024,Li2024c,Li2024b,Xie2023,Pandey2022,Tong2023b,Xia2020a,Tong2023a,Tong2025,Shen2024a}. (d) Normalized $\kappa_c$ ($\kappa_p$/0.2$\times\kappa_c$) versus $\xi_{\rm phDOS}$ for materials with $\kappa_p$ near 0.2 W/mK.(e) Phonon lifetimes of Tl$_3$BSe$_3$ ($\xi_{\rm phDOS}$=0.66). (f) The differential of cumulative $\kappa_p$, $\kappa_c$, and dimensionless phDOS of Tl$_3$BSe$_3$. The inset shows the $\kappa_c$ terms associated with various phonon pairs with frequency ($\omega$, $\omega'$).}
\label{Fig2}
\end{figure*}

Figure~\ref{Fig2}(a) depicts the distinctive phDOS signature of materials capable of simultaneous $\kappa_p$-$\kappa_c$ suppression. The spectrum shows a broad low-frequency peak originating from densely packed phonon branches that enhances phonon scattering to reduce $\kappa_p$, while sparse high-frequency peaks from flat optical modes minimize $\kappa_c$ by suppressing coherent transport channels.

We implemented a systematic screening protocol using the PhononDB database of calculated phonon properties\cite{Togo2015}, as shown in Figure~\ref{Fig2}(b). Our workflow begins with materials containing fewer than 35 atoms per unit cell to ensure manageable computational complexity while maintaining sufficient structural diversity. We then enforce thermodynamic stability by requiring the minimum phonon frequency $\omega_{\rm min}>-0.1$ THz to eliminate dynamically unstable configurations. The screening employs two key descriptors for thermal transport suppression: first, we select materials with maximum acoustic frequencies $\omega_{\rm a-max}<0.8$ THz to ensure low acoustic phonon group velocities that suppress $\kappa_p$; second, we introduce a phDOS sparsity metric $\xi_{\rm phDOS}>0.4$, defined as:

\begin{equation}
\xi_{\rm phDOS} = \frac{\sqrt{N} - \|\tilde{g}\|_1/\|\tilde{g}\|_2}{\sqrt{N} - 1}
\end{equation}

where $\tilde{g}_i \equiv g(\omega_i)/\max[g(\omega)]$ represents the normalized phonon density of states (phDOS), $\|\tilde{g}\|_1$ is the $L_1$ norm, and $\|\tilde{g}\|_2$ is the $L_2$ norm. This dimensionless parameter ranges from 0 for uniform distributions to 1 for maximally sparse single-peak spectra, with benchmark values of $\xi_{\rm phDOS}=0.26$ for MgAgSb and 0.79 for Rb$_6$Re$_6$S$_8$I$_8$.

First-principles calculations identified the top five candidates listed in Table 1, all of which exhibit low values of both $\kappa_p$ and $\kappa_c$. Figure~\ref{Fig2}(c) maps their room-temperature $\kappa_p$ and $\kappa_c$ values against primitive cell atom counts, with screened materials shown as blue circles and reference compounds from both the database and literature as red circles (complete data in Table S2). While conventional materials show the expected trend of decreasing $\kappa_p$ with increasing unit cell complexity (more atoms/unit cell), this typically comes at the cost of enhanced $\kappa_c$ due to reduced phonon band spacing -- a fundamental limitation of traditional approaches to thermal conductivity reduction. The screened materials cluster in the low-conductivity regime, with several candidates achieving total $\kappa_L$ values below 0.25 W/mK. This successful identification validates our descriptor-based strategy for discovering materials with ultralow thermal conductivity while overcoming the conventional $\kappa_p$--$\kappa_c$ trade-off.

To investigate the correlation between $\xi_{\rm phDOS}$ and $\kappa_c$, 
Figure~\ref{Fig2}(d) plots the normalized $\kappa_c$ 
($\kappa_p/0.2 * \kappa_c$) against $\xi_{\rm phDOS}$ for materials with 
$\kappa_p$ near 0.2~W/mK. 
Since $\kappa_p$ and $\kappa_c$ generally exhibit a competitive relationship 
(i.e., materials with lower $\kappa_p$ tend to have higher $\kappa_c$), 
the normalization factor $\kappa_p/0.2$ provides a rough measure of the 
$\kappa_c$ value one would expect if $\kappa_p$ were 0.2~W/mK. 
This allows a fairer comparison of $\kappa_c$ across materials with 
different $\kappa_p$, revealing a clear decreasing trend with increasing 
$\xi_{\rm phDOS}$, which demonstrates that sparse phonon spectra 
(higher $\xi_{\rm phDOS}$) effectively suppress coherent thermal transport.

Figures~\ref{Fig2}(e) and (f) analyze Tl$_3$BSe$_3$, the screened material with the highest $\xi_{\rm phDOS}$ (0.66). The phonon lifetime distribution in Figure~\ref{Fig2}(e) shows that high-frequency modes ($>$2.5 THz) maintain relatively long lifetimes but contribute negligibly to thermal transport. This behavior mirrors that observed in Rb$_6$Re$_6$S$_8$I$_8$, confirming the universal nature of this transport mechanism in materials with high $\xi_{\rm phDOS}$.

The differential thermal conductivity contributions in Figure~\ref{Fig2}(f) reveal that only phonons below 2.5 THz contribute significantly to both $\kappa_p$ and $\kappa_c$ in Tl$_3$BSe$_3$. The $\kappa_p$ spectrum is dominated by acoustic and the lowest optical branches, while $\kappa_c$ originates primarily from interactions between low-frequency modes. The inset shows the $\kappa_c$ terms associated with various phonon pairs with frequency ($\omega$, $\omega'$). Only the phonon pairs near the diagonal in the low-frequency region mainly contribute to $\kappa_c$. These findings confirm that the $\xi_{\rm phDOS}$ metric successfully identifies materials with the desired phonon band engineering -- concentrated low-frequency modes for $\kappa_p$ reduction through enhanced scattering, coupled with sparse high-frequency spectra that minimize $\kappa_c$ contributions. The Tl$_3$BSe$_3$ case study validates our screening approach and reinforces the universal relationship between phonon spectrum sparsity and suppressed coherent thermal transport.

\begin{figure*}[ht!]
\centering
\includegraphics[width=1\linewidth]{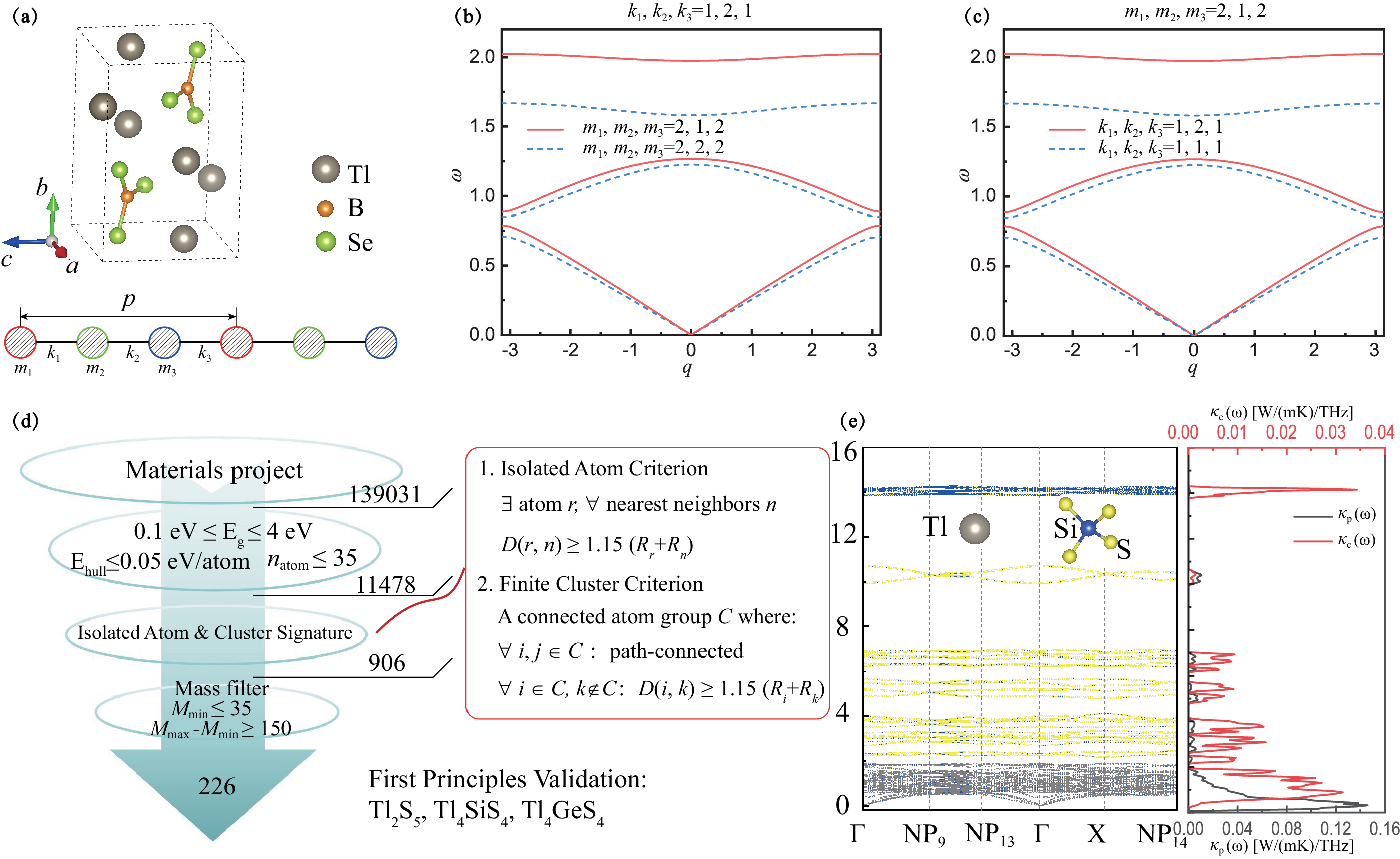}
\caption{(a) Crystal structure of Tl$_3$BSe$_3$. 1D triatomic chain model for materials with co-suppressed $k_p$ and $k_c$. Changes in phonon dispersion of the 1D triatomic chain induced by modifications in (b) atomic mass and (c) elastic constants. (d) Flowchart for high-throughput screening of potential materials with co-suppressed $k_p$ and $k_c$ in the Materials Project using structural descriptors. (e) The projected phonon dispersion of Tl$_4$SiS$_4$ weighted by its constituent atoms. The differential of cumulative $\kappa_p$ and $\kappa_c$.}
\label{Fig3}
\end{figure*}

Structural analysis of the 5 screened materials reveals their common configuration consisting of isolated atoms coupled with clusters, as illustrated by Tl$_3$BSe$_3$ in Fig. 3(a). In this representative structure, Tl atoms interact weakly with the surrounding BSe$_3$ clusters, while the intra-cluster bonds remain strong. This particular arrangement leads to a characteristic ``light-heavy and soft-stiff" hierarchy in the atomic structure, where the mass contrast between constituent atoms and the stiffness difference between inter- and intra-cluster bonds play decisive roles in phonon engineering.

The simplified 1D triatomic chain model [Fig. 3(a)] effectively captures these structural features. As demonstrated in Figs. 3(b) and 3(c), modifications in atomic mass and elastic constants produce distinct effects on the phonon dispersion: (1) Lighter atoms and stronger bonds shift optical branches to higher frequencies, while (2) heavier atoms and weaker bonds predominantly contribute to the acoustic and low-frequency optical branches. This decoupled response suggests an important design principle -- the ``light-heavy and soft-stiff" configuration enables independent control of different phonon regimes, where the heavy/weakly-bonded components govern $\kappa_p$ through enhanced phonon scattering, while the light/strongly-bonded units suppress $\kappa_c$ by creating spectral sparsity.

Building upon the structural insights from Figure~3(a)--(c), 
we implement a multi-step screening protocol to identify materials 
with the desired isolated-atom/cluster configuration. 
The workflow begins with three pre-screening criteria applied to 
the Materials Project database\cite{Jain2013}. 
We require band gaps $E_g$ between $0.1$--$4$~eV to ensure semiconducting behavior. 
The energy above hull is restricted to $E_{hull} \leq 0.05$~eV/atom 
to guarantee thermodynamic stability. 
The primitive cell atom count is limited to $n \leq 35$ 
to control computational complexity. 
Our structural descriptor then isolates candidates exhibiting 
both isolated atoms and clustered units. 
Isolated atoms are defined as atoms with no chemical bonds to neighbors, 
where $D(r,n)$ is the interatomic distance between atoms $r$ and $n$, 
$R_r$ and $R_n$ are their covalent radii, 
and $D(r,n) \geq 1.15(R_r + R_n)$ indicates the absence of a chemical bond. 
Clustered units are internally connected atomic groups with no external bonds. 
A mass contrast requirement is also applied 
($M_{\text{min}} \leq 35$ and $\Delta M = M_{\text{max}} - M_{\text{min}} \geq 150$) 
to ensure phonon band separation. 
This approach yields $226$ qualified materials, 
including three Tl--S-based compounds -- Tl$_2$S$_5$, Tl$_4$SiS$_4$, 
and Tl$_4$GeS$_4$ -- that exhibit simultaneous suppression of both 
$\kappa_p$ and $\kappa_c$ as listed in Table~1.

The projected phonon spectrum of Tl$_4$SiS$_4$ in Figure~3(e) reveals the characteristic phonon engineering of these materials. The structure consists of isolated Tl atoms ionically bonded to rigid SiS$_4$ clusters, creating distinct vibrational contributions: (i) The light Si atoms, strongly covalently bonded to S, dominate the high-frequency optical modes near $14$~THz; (ii) Intermediate-frequency modes ($2$--$11$~THz) primarily arise from S vibrations; (iii) The heavy Tl atoms, weakly ionically coupled to the clusters, generate densely packed acoustic and low-frequency optical branches below $2$~THz. This hierarchical spectral distribution -- dense low-frequency modes from Tl for $\kappa_p$ suppression and sparse high-frequency modes from SiS$_4$ for $\kappa_c$ reduction -- exemplifies the "heavy-light, soft-stiff" design principle demonstrated in Figure~3(b,c). It can also be seen from the differential of cumulative $\kappa_c$ that the high-frequency optical branches exhibit relatively small contribution to $\kappa_c$.

\begin{table*}
\caption{Selected ultralow thermal conductivity materials with their chemical formulas, space group symmetries, $\kappa_p$ (W/mK), $\kappa_c$ (W/mK), $\xi_{\rm phDOS}$, and number of atoms per primitive cell $n$.}
\begin{ruledtabular}
\begin{tabular}{cccccc}
 materials&symmetry&$\kappa_p$&$\kappa_c$
&$\xi_{\rm phDOS}$&$n$\\ \hline
 Cs$_3$PSe$_4$&$Pnma$&0.22&0.06&0.62&32 \\
 NaAlI$_4$&$Pnma$&0.07&0.07&0.56&24\\
 Rb$_4$Au$_6$S$_5$&$P\bar{6}2c$&0.16&0.06&0.59&30\\
 Tl$_3$PSe$_4$&$Pnma$&0.18&0.06&0.53&32\\
 Tl$_3$BSe$_3$& $P2_1/m$&0.20&0.05&0.66&14\\
 Tl$_2$S$_5$& $P2_{1}2_{1}2_{1}$ &0.17&0.08&0.58&28\\
 Tl$_4$SiS$_4$& $Cc$ &0.10&0.06&0.55&18\\
 Tl$_4$GeS$_4$& $Cc$ &0.09&0.06&0.48&18\\
\end{tabular}
\end{ruledtabular}
\end{table*}

In summary, this work advances the fundamental understanding of phonon engineering by establishing three key insights: (1) The "heavy-light and soft-stiff" design principle represents a universal strategy to decouple $\kappa_p$ and $\kappa_c$ suppression mechanisms, overcoming their traditional trade-off; (2) The hierarchical phonon spectrum emerging from this configuration -- with heavy/soft components governing low-frequency scattering and light/stiff units controlling high-frequency coherence -- provides a new paradigm for thermal transport modulation; (3) The success of our structural descriptor, validated through theoretical modeling and first-principles calculations, demonstrates that targeted phonon band engineering is achievable through careful atomic-scale design. These findings not only resolve long-standing challenges in thermal conductivity reduction but also open avenues for discovering materials with precisely tailored phonon properties. 

\section*{Acknowledgements}

This work is supported by the Natural Science Foundation of China (Grants No. 12304038, 52206092, 12204402), the National Key Research and Development Program of China (Grants No. 2024YFA1409800), the Big Data Computing Center of Southeast University and the Center for Fundamental and Interdisciplinary Sciences of Southeast University.

%


\end{document}